\documentclass[aps,prc,amsmath,amssymb,twocolumn,aps,showpacs,superscriptaddress,nofootinbib]{revtex4-1}

\usepackage{graphicx,amssymb,amsmath}
\usepackage[dvips]{xcolor}
\usepackage{hyperref}
\usepackage{amsmath}
\usepackage{amssymb}

\usepackage{soul}

\newcommand{\be}{\begin{equation}}
\newcommand{\ee}{\end{equation}}
\newcommand{\ba}{\begin{eqnarray}}
\newcommand{\ea}{\end{eqnarray}}

\begin{document}
\title{Mean field and two-body nuclear effects in inclusive electron scattering
on argon, carbon and titanium: the superscaling approach}
 \author{M.B.~Barbaro}
 \affiliation{Dipartimento di Fisica, Universit\`{a} di Torino and INFN, Sezione di Torino, Via P. Giuria 1, 10125 Torino, Italy}
 \author{J.A.~Caballero}
 \affiliation{Departamento de F\'{i}sica At\'omica, Molecular y Nuclear, Universidad de Sevilla, 41080 Sevilla, Spain}
 \author{A.~De Pace}
 \affiliation{INFN, Sezione di Torino, Via P. Giuria 1, 10125 Torino, Italy}
\author{T.W.~Donnelly}
\affiliation{Center for Theoretical Physics, Laboratory for Nuclear Science and Department of Physics, Massachusetts Institute of Technology, Cambridge, Massachusetts 02139, USA}
\author{R.~Gonz\'alez-Jim\'enez}
\affiliation{Grupo de F\'isica Nuclear, Departamento de Estructura de la Materia, F\'isica T\'ermica y Electr\'onica, Facultad de Ciencias F\'isicas, Universidad Complutense de Madrid and IPARCOS, Madrid 28040, Spain}
\author{G.D.~Megias}
\affiliation{Departamento de F\'{i}sica At\'omica, Molecular y Nuclear, Universidad de Sevilla, 41080 Sevilla, Spain}
\affiliation{IRFU, CEA, Universit\'e Paris-Saclay, 91191 Gif-sur-Yvette, France}

\date{\today}
\begin{abstract}
We compare the predictions of the SuSAv2 model including two-particle two-hole meson-exchange currents with the recent JLab data for inclusive electron scattering on three different targets (C, Ar and Ti). The agreement is very good over the full energy spectrum, with some discrepancy seen only in the deep inelastic region. The 2p2h response, peaked in the dip region between the quasielastic and $\Delta$-resonance peak, is essential to reproduce the data. 
We also analyze the $k_F$ (Fermi momentum) dependence of the data in terms of scaling of second kind, showing that the 2p2h response scales very differently from the quasielastic one, in full accord with what is predicted by the model. 
The results represent a valuable test of the applicability of the model to neutrino scattering processes on different nuclei. 
\end{abstract}


\maketitle


The precise description of electron scattering data on complex nuclei is not only interesting by itself, but also of crucial importance in connection with current and future neutrino oscillation experiments \cite{DUNE16,HyperK15,T2K11,NOvA07,MiniBooNE09,MicroBooNE12,PhysRevLett.121.022504,PhysRevD.98.032003}, which aim at measuring the leptonic CP violation phase, $\delta_{CP}$, assessing the neutrino mass hierarchy and improving the precision on the oscillation mixing angles. 
A few percent uncertainty in the description of neutrino-nucleus interactions occurring in the detectors of these experiments, typically composed of argon, water or mineral oil, is needed in order to meet the required precision in the experimental analyses (see \cite{Alvarez-Ruso17,Katori17} for recent reviews of this subject).
The importance of validating different nuclear models using high quality electron scattering data in the relevant energy domain 
has been often stressed \cite{Meucci09,Gallmeister16,Pandey15,Ivanov16,Rocco16,Lovato16} and careful comparisons have been performed with the carbon and oxygen data collected in the QES archive \cite{QES}.
The recent  measurement of Ar$(e,e')$X and  Ti$(e,e')$X cross sections performed at Jefferson Lab \cite{Dai:2018xhi,Dai:2018gch} offers a new opportunity to test nuclear models~\cite{Ankowski:2005wi,Butkevich:2012zr,Gallmeister:2016dnq,VanDessel:2017ery} in a kinematic region and on nuclei that are specifically relevant for neutrino oscillation experiments.

In  \cite{Amaro:2004bs} it was first suggested that the superscaling properties of inclusive $(e,e')$ data represent a powerful tool to connect electron- and neutrino-scattering reactions.
Detailed studies of scaling and superscaling for electron-nucleus cross sections
have been presented in \cite{Donnelly99a,Donnelly99b,Maieron:2001it}. The analysis of
the $(e,e')$ world data has shown the quality of the scaling
behavior: scaling of the first kind (no dependence
on the momentum transfer) is quite good at
excitation energies below the QE peak, whereas scaling of second
kind (no dependence on the nuclear species) works extremely well in the
same region.

The Relativistic Mean Field (RMF) model is able to reproduce with good accuracy the superscaling properties of the data and gives a good microscopic description of quasi-elastic electron scattering data~\cite{Gonzalez-Jimenez14b}.
The RMF model describes the bound state nucleons as single-particle wave functions obtained by solving the Dirac equation with self-consistent relativistic mean-field potentials~\cite{Serot97,Horowitz81,Sharma93}. 
The outgoing nucleon is described as a continuum wave function, specifically, the solution of the Dirac equation with the same strong mean-field potentials used to describe the bound state. This implies that final-state interactions (FSI) are taken into account within a fully relativistic framework and the model is free from 
non-orthogonality effects. As  shown in \cite{Meucci09,Maieron03,Gonzalez-Jimenez14b}, RMF predictions compare remarkably well with inclusive electron scattering data up to moderated $q$-values, {\it i.e.,} up to values of the knock-out nucleon kinetic energy, $T_N$, of the order of 150 MeV. 
For larger values, {\it i.e.,} high momentum transfer, the energy-independent scalar and vector potentials become    
unrealistically strong, leading to results that depart very significantly from data~\cite{Gonzalez-Jimenez14b}. On the contrary, the relativistic plane wave impulse approximation (RPWIA), where FSI are turned off, yields better predictions. Notice that at these high nucleon momenta ($\geq 500$ MeV) the overlap between initial and final states is sufficiently small to prevent spurious contributions in the cross section.

The SuSAv2 model~\cite{Gonzalez-Jimenez14b,Megias16a} takes advantage of these observations and, on the base of superscaling~\cite{Donnelly99a,Donnelly99b}, builds an effective model that incorporates both regimes, RMF and RPWIA, making use of a ``blending'' function. 
The model extended to the inelastic spectrum and including the important contribution of two-particle-two-hole (2p2h) excitations induced by Meson Exchange Currents (MEC) \cite{DePace:2003spn,DePace:2004cr}, provides amazingly good agreement with $^{12}$C$(e,e')$ and $^{16}$O$(e,e')$ data for very different kinematics and covering almost the whole range of energy transfer~\cite{Megias16a}. Only
at very low $q$ and $\omega$ values does one find that SuSAv2 fails, and this should be expected since this is the region where scaling breaks and collective nuclear effects are important. Likewise, the SuSAv2-MEC model has also demonstrated a good level of compromise with recent measurements of neutrino and antineutrino reactions on carbon and oxygen~\cite{Megias:2016nu,Megias:2018oxygen,Megias:2014kia,Megias:2018minervanubar}.

In this work we present and discuss the SuSAv2-MEC predictions with the new JLab data. We also perform an analysis of the nuclear dynamics and the scaling phenomenon for the different nuclear targets involved by means of the RMF and RPWIA approaches. Finally we study the superscaling behavior of the JLab data both in the quasi-elastic and in the 2p2h regimes to stress the validity of the SuSAv2-MEC model.




\begin{figure*}[ht]   \vspace*{-0.495cm}        
\includegraphics[scale=0.209, angle=270]{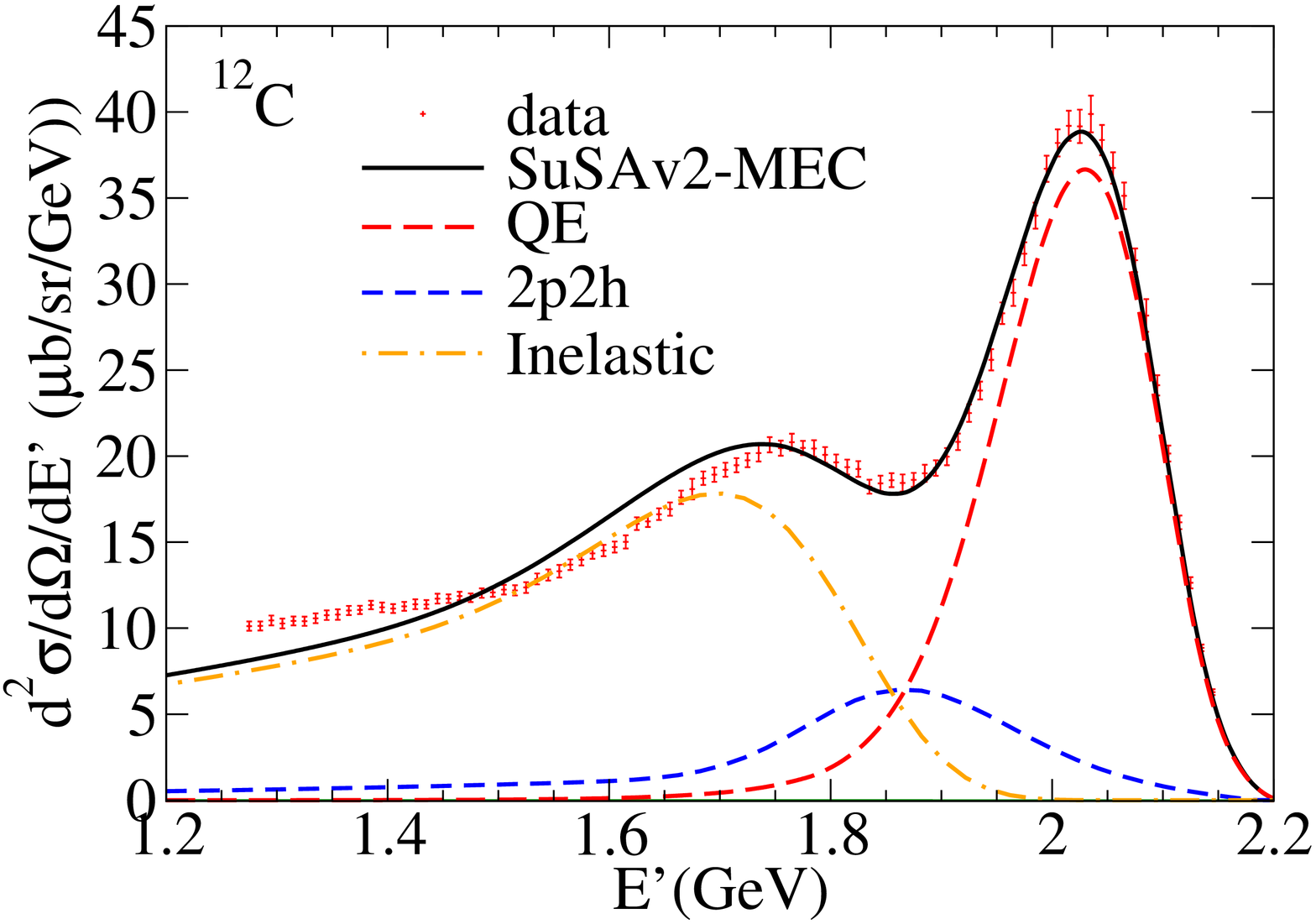}\hspace*{-0.39cm} 
\includegraphics[scale=0.209, angle=270]{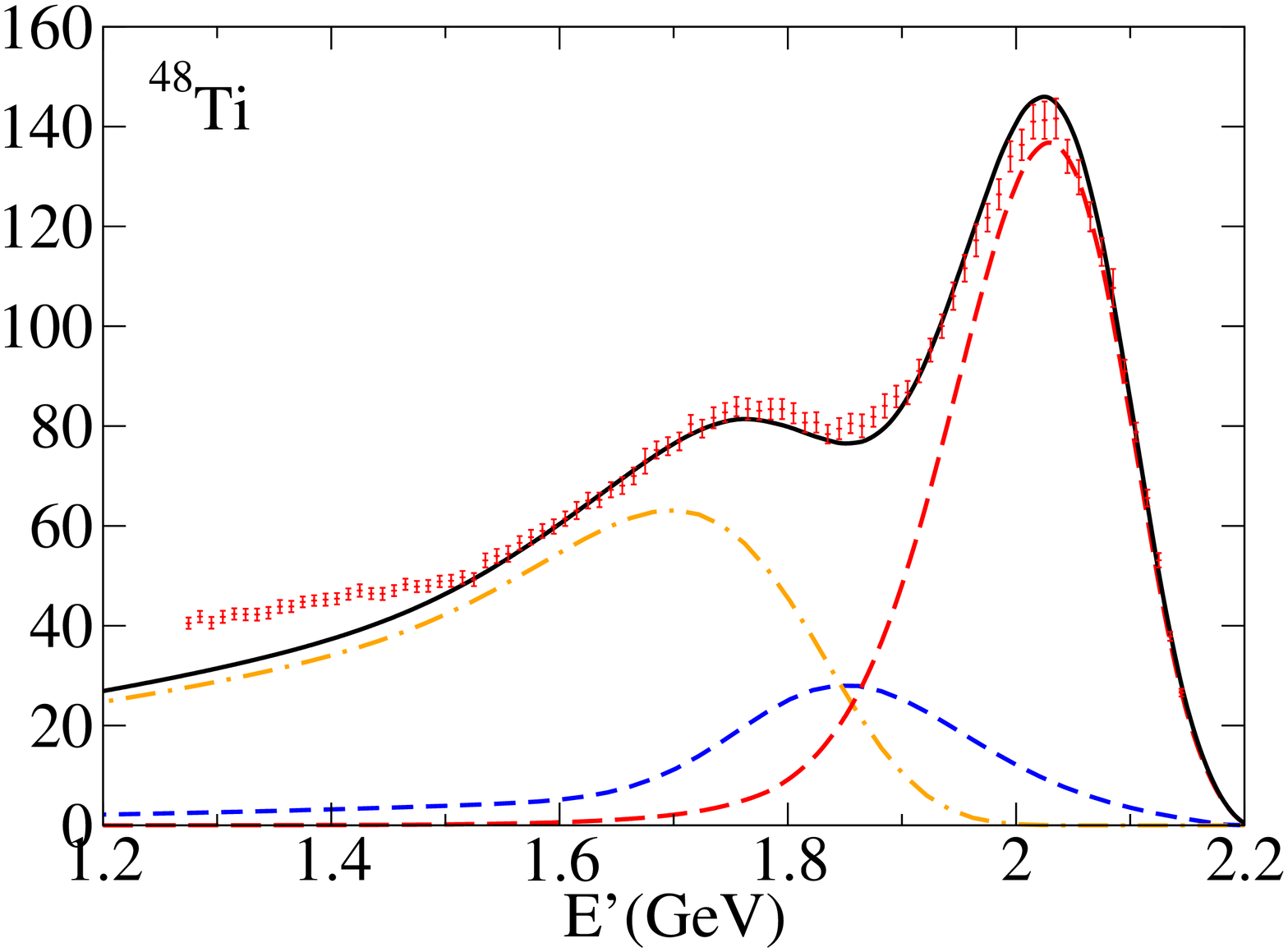}\hspace*{-0.39cm} 
\includegraphics[scale=0.209, angle=270]{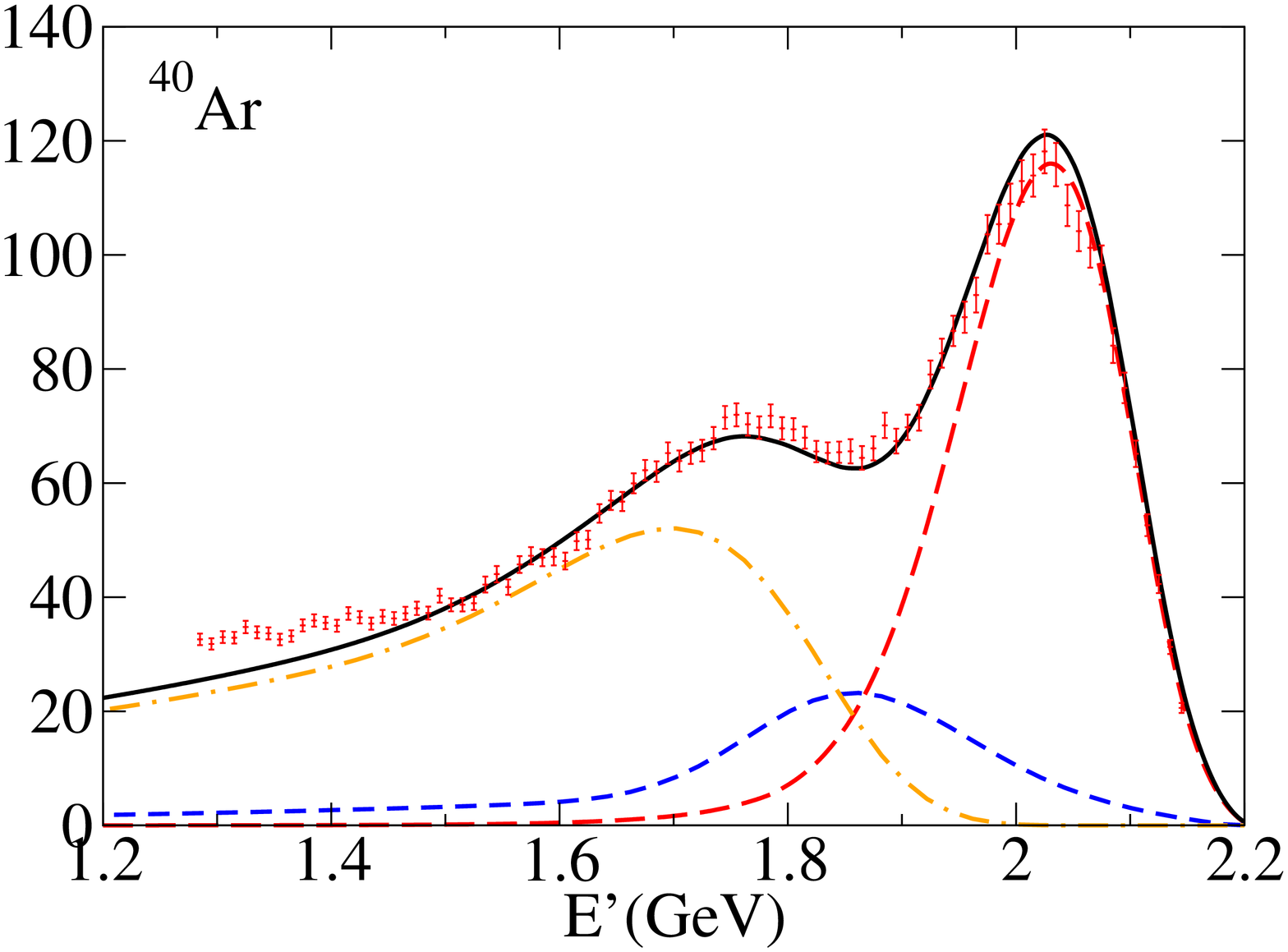}\\\vspace*{-0.695cm}
\includegraphics[scale=0.209, angle=270]{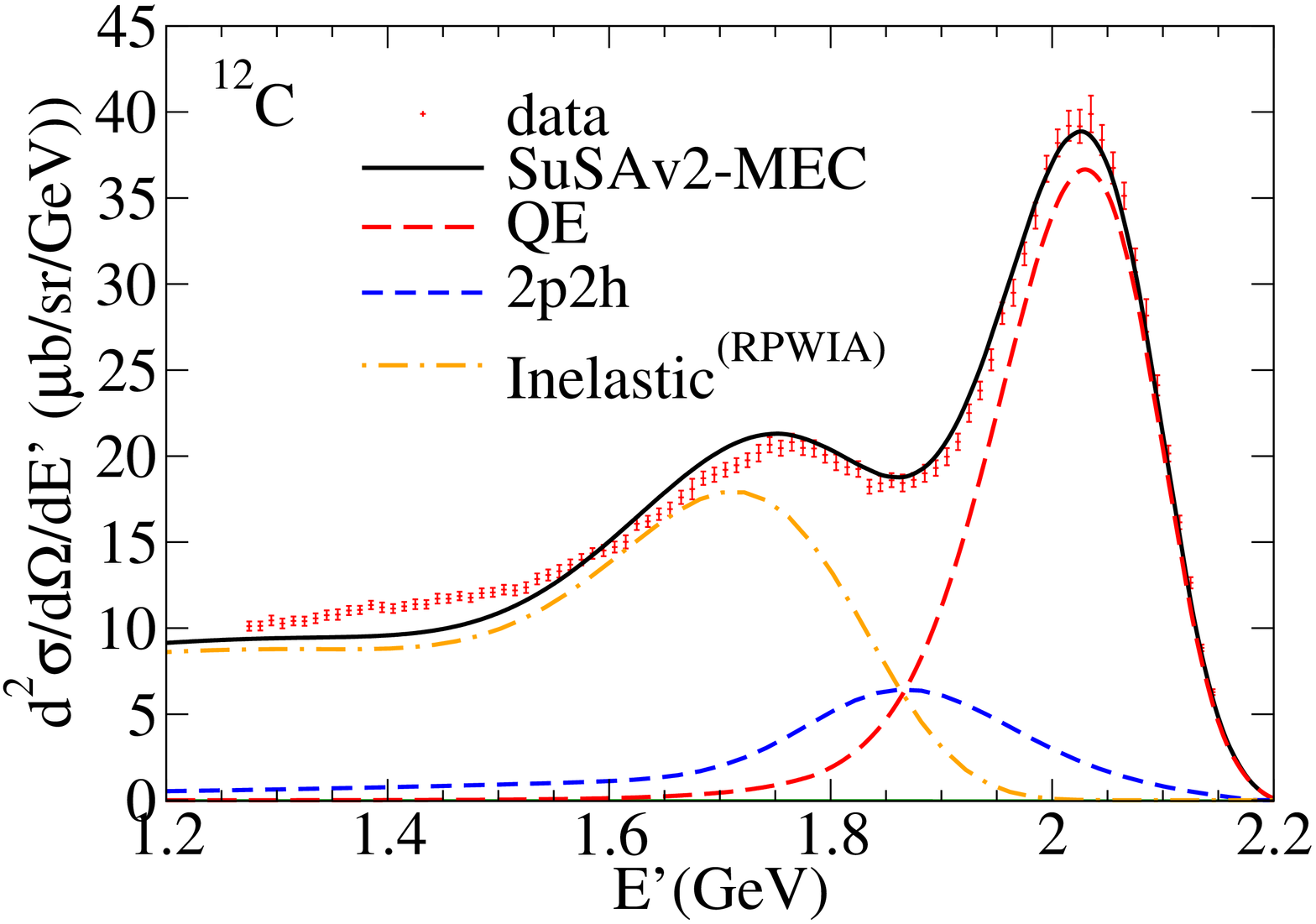}\hspace*{-0.39cm} 
\includegraphics[scale=0.209, angle=270]{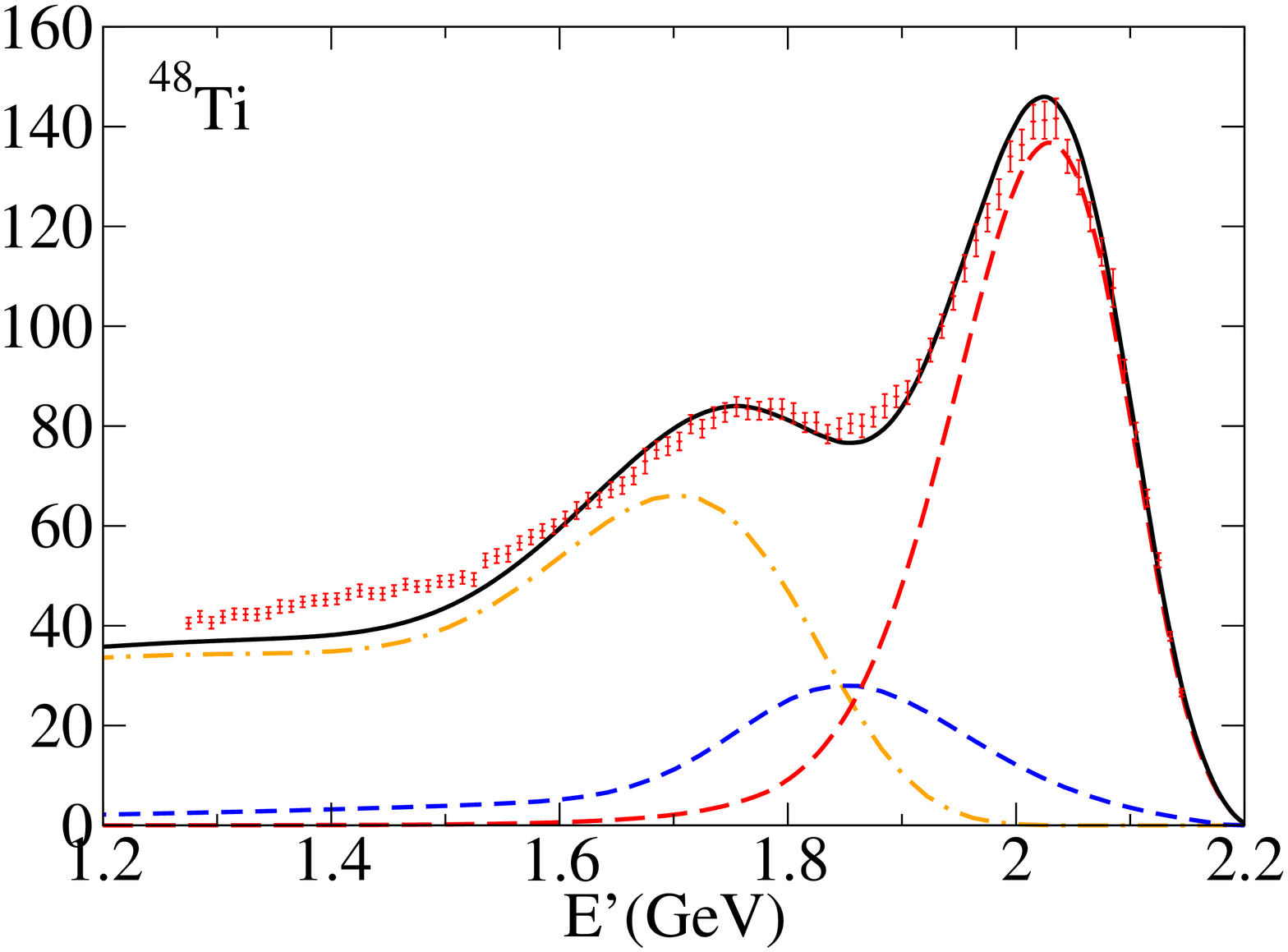}\hspace*{-0.39cm} 
\includegraphics[scale=0.209, angle=270]{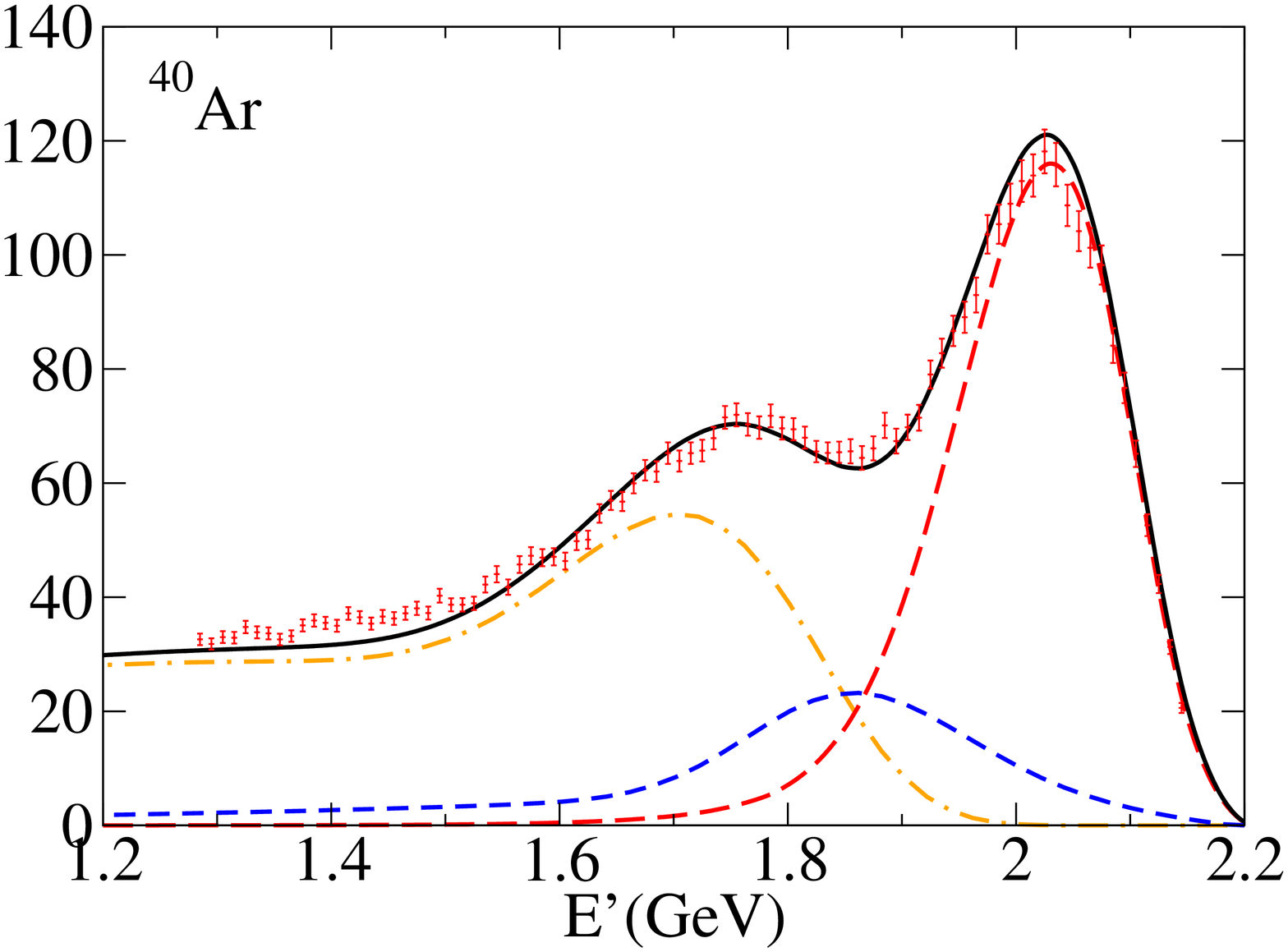}\vspace*{-0.195cm}
		\caption{(Color online)  Top panels: The $(e,e')$ double differential cross section of carbon, titanium and argon  from  \cite{Dai:2018xhi,Dai:2018gch}, compared with the SuSAv2-MEC prediction. For completeness, the separate QE, 2p2h and inelastic contributions are also shown. The beam energy is $E$=2.222 GeV and the scattering angle $\theta$=15.541 deg. Bottom panels: As top panels, but using the RPWIA model for the inelastic cross section. }\label{fig:fig1}
\end{figure*}

In Fig.~\ref{fig:fig1} we present the predictions of the SuSAv2-MEC model compared with data for carbon (left panel), titanium (middle) and argon (right). The values used for the Fermi momentum, $k_F$, are 0.228 GeV/c for carbon and 0.241 GeV/c for titanium and argon. 
As also addressed in previous studies~\cite{Megias16a,Maieron:2001it}, a shift in the transferred energy has also been included in all cases. Moreover, the extension of the SuSAv2 model to other nuclei, both symmetric and asymmetric, beyond $^{12}$C is based as well on RMF and RPWIA predictions~\cite{Gonzalez-Jimenez14b} but considering the neutron and proton form factors weighted by the corresponding neutron (N) and proton (Z) numbers. The differences between neutrino reactions on symmetric and asymmetric nuclear targets were carefully studied in~\cite{Megias18} within the Relativistic Fermi Gas (RFG) framework. Following this line, a dedicated analysis of the SuSAv2 predictions in asymmetric nuclei will be shortly presented~\cite{inpreparation1}.
For the QE regime, we describe the electromagnetic nucleon form factors using the extended Gari-Krumpelmann (GKex) model~\cite{GKex}. 
The sensitivity of the responses to other parametrizations has been discussed in \cite{Megias13}. With regard to the inelastic structure functions, we adopt the Bosted-Christy parametrization~\cite{Bosted08,Christy10} that provides a good description of the resonant structures in $(e,e')$ cross sections covering a wide kinematic region. As shown in \cite {Megias:2017PhD}, the use of other choices such as the Bodek-Ritchie~\cite{Bodek81} parametrization and models based on Parton Distribution Functions (PDF) leads to very large discrepancies with $(e,e')$ data.

In each graph the separate contributions corresponding to the QE (red dashed line), the two-particle two-hole MEC (blue short-dashed) and inelastic (green dot-dashed) channels are shown. The total contribution is represented by the solid line. The excellent agreement between theory and data is excellent over most of the energy spectrum covering the QE, dip and a significant region in the inelastic domain. Only at the largest values of the transferred energy (smallest values for the ejected electron energy) do the theoretical predictions depart from data, these being larger. 

The origin of this discrepancy is related to the specific procedure used to determine the RMF/RPWIA transition in the SuSAv2 model in the inelastic regime. %
Details on this analysis as well as a study on the sensitivity of the $(e,e')$ cross sections to different choices of the parameters were given in \cite{Megias16a}. In Fig.~\ref{fig:fig1} we use the model as presented in \cite{Megias16a}, even being aware that better agreement with the high inelastic data could be achieved by employing other options for the transition parameters. 
To illustrate this point, we show in the bottom panels of Fig.~\ref{fig:fig1} the predictions of a modified model where the inelastic region is treated ignoring final-state interactions, namely keeping only the RPWIA component of the SuSAv2 model. It appears that the flat behavior of the cross section at low $E^\prime$ is better reproduced by the pure RPWIA inelastic contribution, indicating that FSI are somewhat 
overestimated at very high energy in the present SuSAv2-inelastic model. An updated version of the model, which takes into account also these new JLab data in order to improve description of the inelastic regime at very high kinematics, will be presented in a forthcoming publication~\cite{inpreparation2}, the focus of the present work being more on the QE and dip regions.

The level of accordance between data and theory in the dip region where similar contributions emerge from the three domains, {\it viz.} QE, 2p2h-MEC and inelastic, is also outstanding.  This can be considered to be a crucial test for the validity of the model, and particularly, the description of the 2p2h-MEC contribution that reaches its maximum value in this region. Comparison between theory and data shows excellent agreement for the three nuclei, being even better in the case of the two heavier asymmetric systems, $^{48}$Ti and $^{40}$Ar. This is reassuring as this is the first time the SuSAv2-MEC model is applied to nuclei with isospin different from zero. It is important to point out that SuSAv2 is entirely based on results obtained with the Relativistic Mean Field (RMF) model and how well they fulfill scaling arguments. To make contact with results in Fig.~\ref{fig:fig1}, in what follows we discuss in detail RMF but applied for the first time to asymmetric nuclei. This is crucial to reinforce our confidence in the validity of the SuSAv2 approach extended to nonzero isospin nuclei.
\begin{figure*}[ht]
\includegraphics[scale=0.223, angle=270]{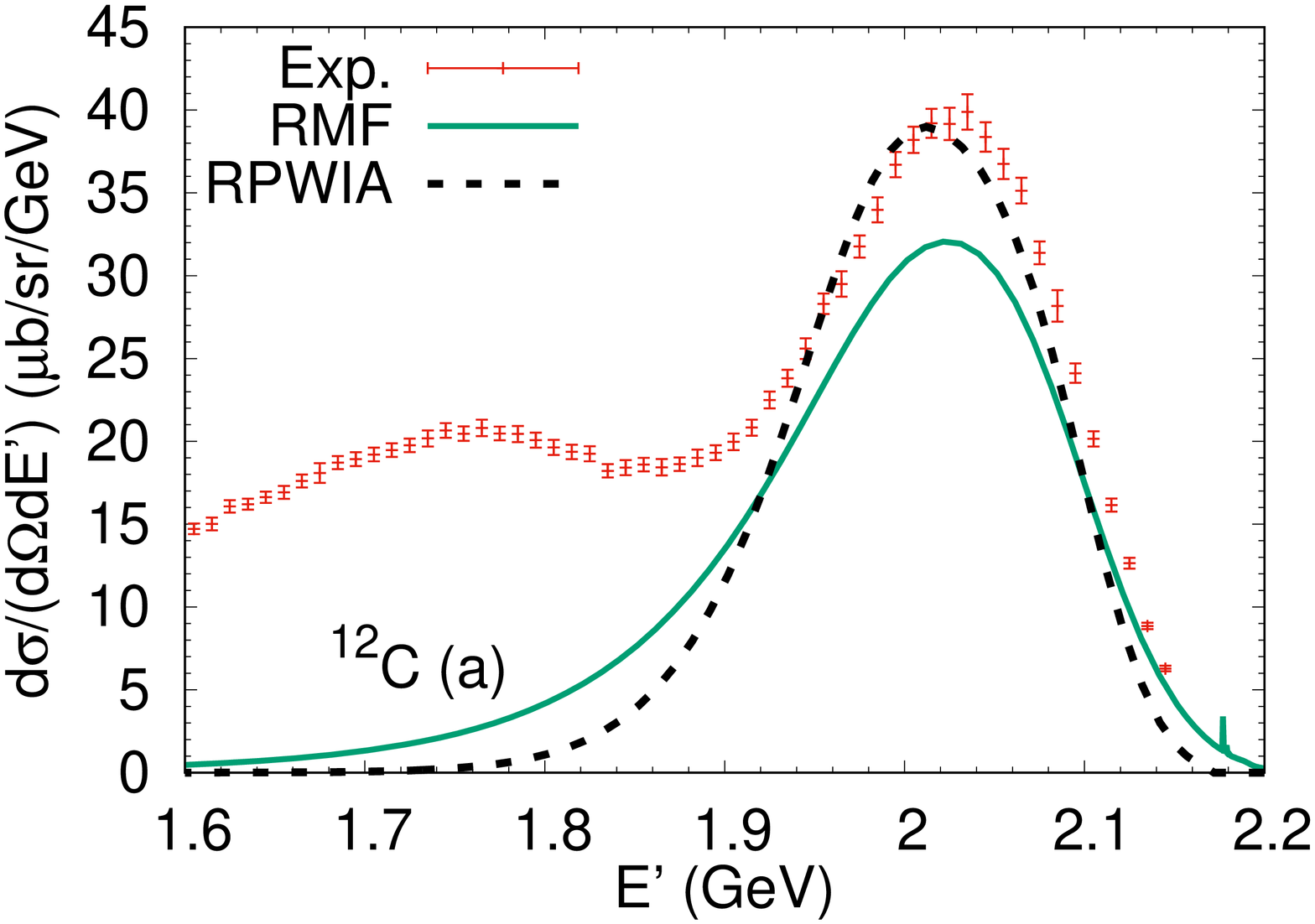} 
\includegraphics[scale=0.223, angle=270]{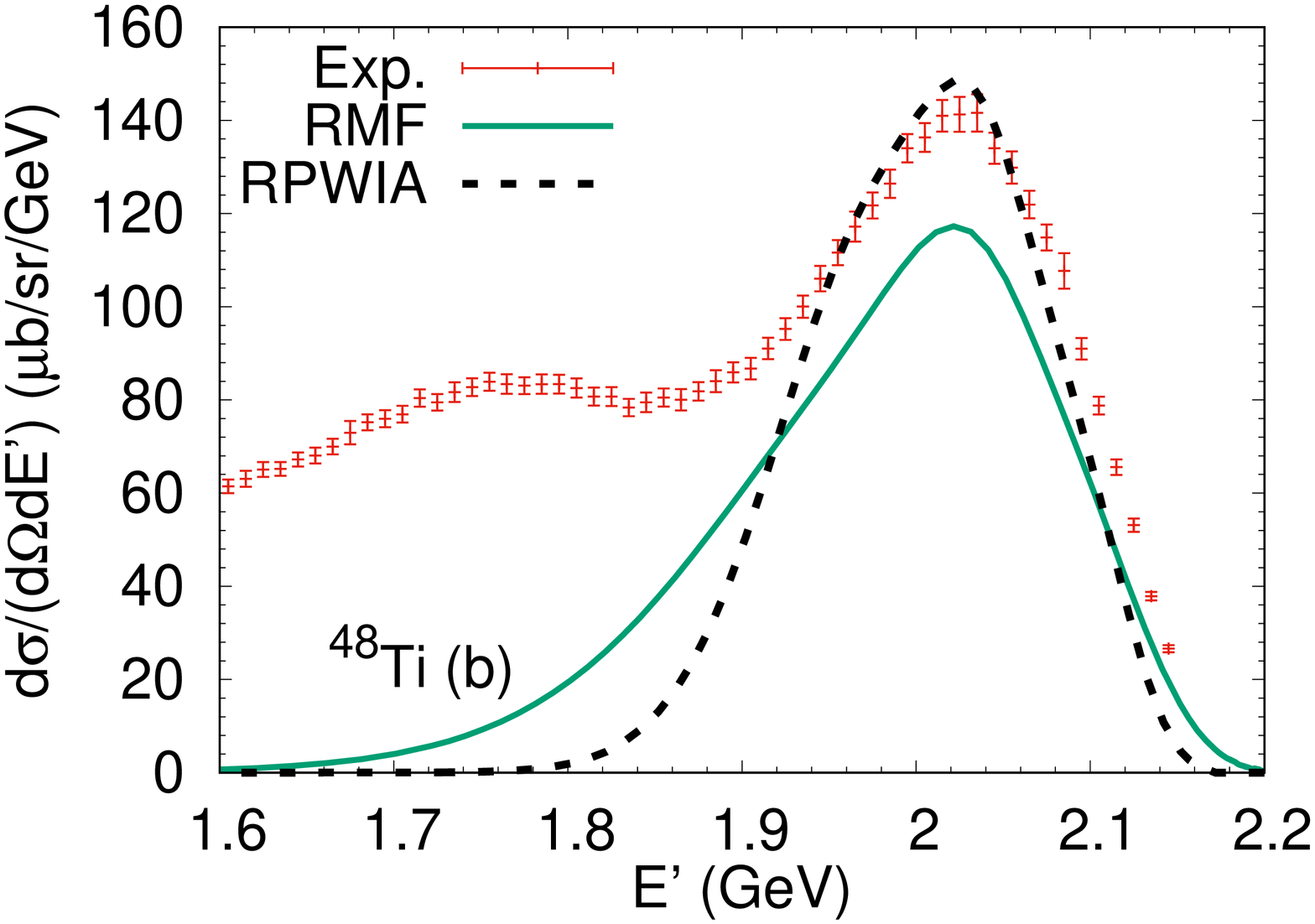}
\includegraphics[scale=0.223, angle=270]{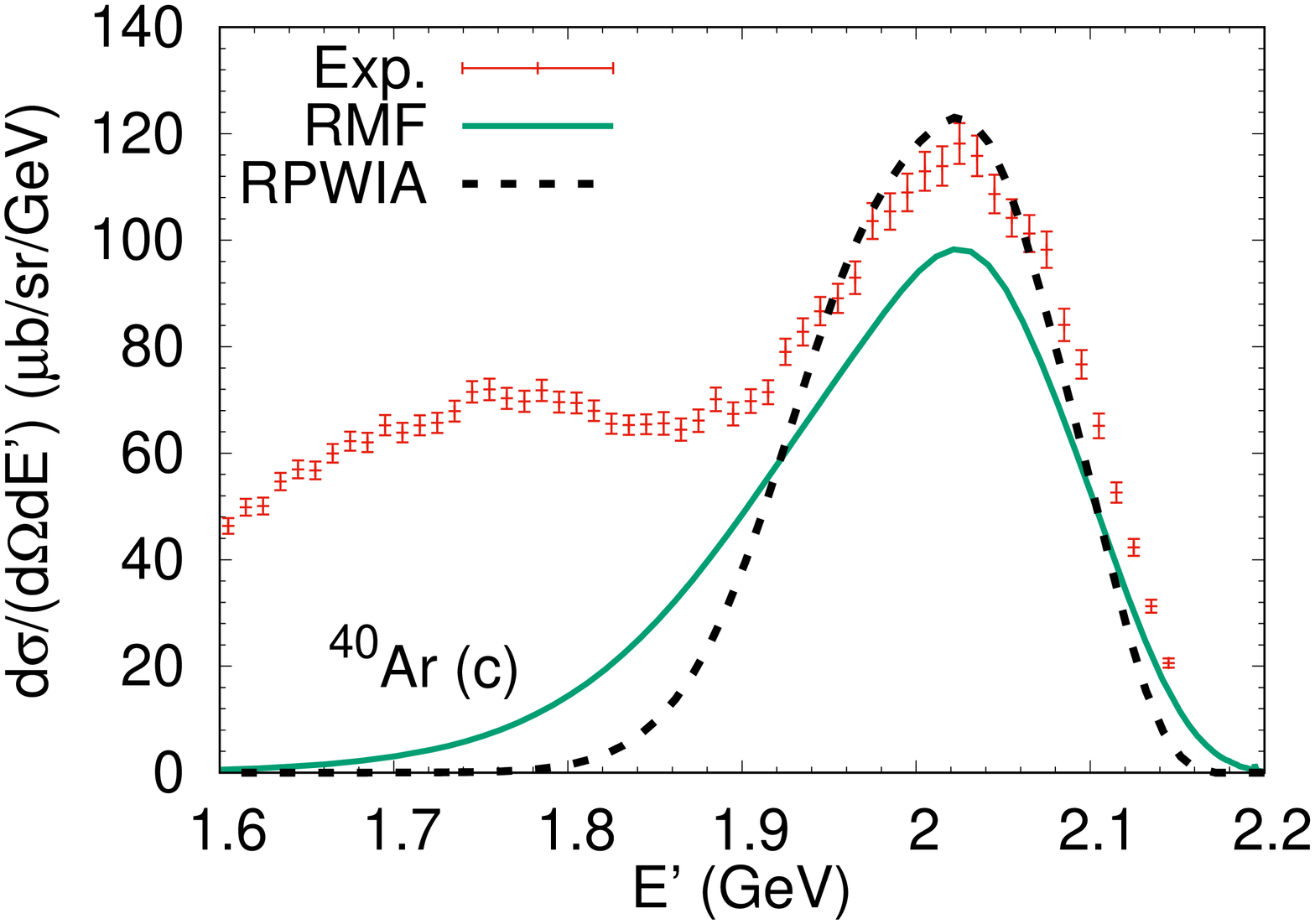}\vspace*{-0.15cm}
\caption{(Color online) The $(e,e')$ double differential cross for $^{12}$C (a), $^{40}$Ar (b), and $^{48}$Ti (c) target nuclei. The theoretical predictions for the quasielastic response are computed with RMF and RPWIA approaches.}
\label{fig:RMF-RPWIA}
\end{figure*}

In Fig.~\ref{fig:RMF-RPWIA} we restrict the comparison to the QE regime. The solid line represents the RMF prediction, {\it i.e.,} with FSI incorporated through the scalar and vector potentials in the final state, whereas the dashed line refers to the RPWIA result.  
As observed, the general behavior of the curves for the two asymmetric nuclei, $^{48}$Ti and $^{40}$Ar, is similar to the 
case of $^{12}$C. In the three cases, the RMF provides a cross section with a long tail extending to smaller values of the scattered electron energy (larger energy transferred). On the contrary, the RPWIA shows a much more symmetrical shape but with the maximum being higher and more in accordance with data (slightly overestimating them). The values of the energy and momentum transfer in the QE peak are sufficiently large ($q\approx600$ MeV) to explain why in the SuSAv2 the RPWIA contribution becomes more important, being the main responsible for the QE response (see \cite{Gonzalez-Jimenez14b,Megias16a} for details).

In what follows we investigate the validity of scaling arguments within the
RMF/RPWIA model, but including in the analysis cross sections corresponding to non-zero isospin nuclei, namely, titanium and argon. Note that this is the basis of the SuSAv2 model. To make the discussions clearer, we start by presenting the basic expressions needed to get the scaling function. 
For inclusive QE electron scattering processes, the superscaling function 
is evaluated by dividing the differential cross section by the
appropriate single-nucleon $eN$ elastic cross section weighted by the corresponding proton and
neutron numbers~\cite{Donnelly99a,Donnelly99b,Maieron:2001it,Barbaro:1998gu} involved in the process:
\be
f(\psi',q)\equiv k_F\frac{\left[\displaystyle\frac{d\sigma}{d\varepsilon'd\Omega'}\right]_{(e,e')}}
{\sigma_M\left[V_LG_L(q,\omega)+V_TG_T(q,\omega)\right]} \, ,
\label{fscaling}
\ee
where we have introduced the dimensionless scaling variable denoted as $\psi'(q,\omega)$,
\be
\psi^\prime\equiv\frac{1}{\sqrt{\xi_F}}\frac{\lambda^\prime-\tau^\prime}
 {\sqrt{(1+\lambda^\prime)\tau^\prime+
\kappa\sqrt{\tau^\prime(1+\tau^\prime)}}} \, 
\label{eq11}
\ee
with $\lambda^\prime\equiv (\omega-E_{shift})/2m_N$,
$\kappa\equiv q/2m_N$, $\tau^\prime\equiv \kappa^2-\lambda^{\prime
2}$, and $\xi_F\equiv\sqrt{1+(k_F/m_N)^2}-1$. 
The term $k_F$ is the Fermi momentum and the energy shift
$E_{shift}$, taken from~\cite{Maieron:2001it}, has been introduced to force the maximum of the cross section to occur for $\psi'=0$. As usual the notation $\psi$ refers to the scaling variable when $E_{shift}=0$.
The single-nucleon functions $G_L$ and $G_T$ are given by\footnote{Here we retain only the lowest-order terms of $G_L$ and $G_T$ in powers of $\eta_F\equiv k_F/m_N$ (see \cite{Amaro:2004bs,PhysRevC.71.065501} for details)}
\be
G_L =\frac{\kappa\widetilde{G}^2_E}
{2\tau} \, , \,  \, \, \,
G_T = \frac{\tau\widetilde{G}^2_M}
{\kappa} \label{gt}\, .
\ee
As usual one has
\ba
&&\widetilde{G}_E^2\equiv ZG_{Ep}^2+NG_{En}^2\, , \,\,\,\,\, 
\widetilde{G}_M^2\equiv ZG_{Mp}^2+NG_{Mn}^2\, ,
\label{gsquare}
\ea
involving the proton and neutron form factors weighted by the proton and neutron numbers $Z$ and $N$, respectively. 

\begin{figure}[h]
\includegraphics[scale=0.25, angle=270]{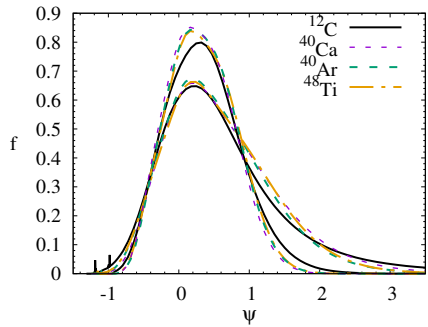}\vspace*{-0.1cm}
\caption{(Color online) Superscaling function $f(\psi)$ evaluated with the RMF and RPWIA models for the following nuclear systems: $^{12}$C, $^{40}$Ca, $^{40}$Ar, and $^{48}$Ti.}
\label{fig:fscaling}
\end{figure}

At sufficiently high energies the function $f$ extracted from the data depends only on the scaling variable
$\psi$ but not on the transferred momentum $q$. Moreover, $f(\psi)$ also becomes independent of
the momentum scale in the problem, that is, independent of $k_F$.  In the context of the RMF and RWIA approaches, this is illustrated in Fig.~\ref{fig:fscaling} where we present $f(\psi)$ for $^{12}$C, $^{40}$Ca, $^{48}$Ti and $^{40}$Ar.  The values of the Fermi momentum for carbon, titanium and argon are the ones mentioned before, while $k_F$ for Ca has been set to 241 MeV/c. Note the different shape in the scaling function when comparing RMF and RPWIA, the former with a long tail extended to high values of $\psi$ (large values of the transfer energy), whereas the latter presents a much more symmetrical shape with the maximum being significantly higher. In both models, the scaling function is very similar for all of the nuclei, particularly for the three heavier ones, Ca, Ti and Ar. Only the case of carbon departs a little bit in the tail (RMF) or in the maximum (RPWIA). This is partly connected with the very different bound nucleon shells involved in the various calculations. In spite of that, scaling of the second kind, {\it i.e.,} independence on the nuclear system, is fulfilled to high precision in all of the cases, making it possible to define a general scaling function to be applicable also for asymmetric nuclei. This is the basis of the SuSAv2 model that here is applied for the first time to nuclei with different number of protons and neutrons providing results in excellent agreement with data (see Fig.\ref{fig:fig1}).
  
 
 Next we perform an analysis of the superscaling behavior of the JLab data not only in the quasielastic region, but also in the ``dip" region between the QE  and $\Delta$ peaks.  We recall that this domain is particularly relevant for the analyses of neutrino oscillation experiments, where, due to the broadly distributed neutrino flux, the 2p2h response is part of the so-called ``QE-like" or ``QE-0$\pi$" cross section and cannot be disentangled from the genuine QE response.
 

\begin{figure*}[ht]\vspace*{-0.2cm}
\hspace*{-0.14cm}\includegraphics[scale=0.67, angle=0]{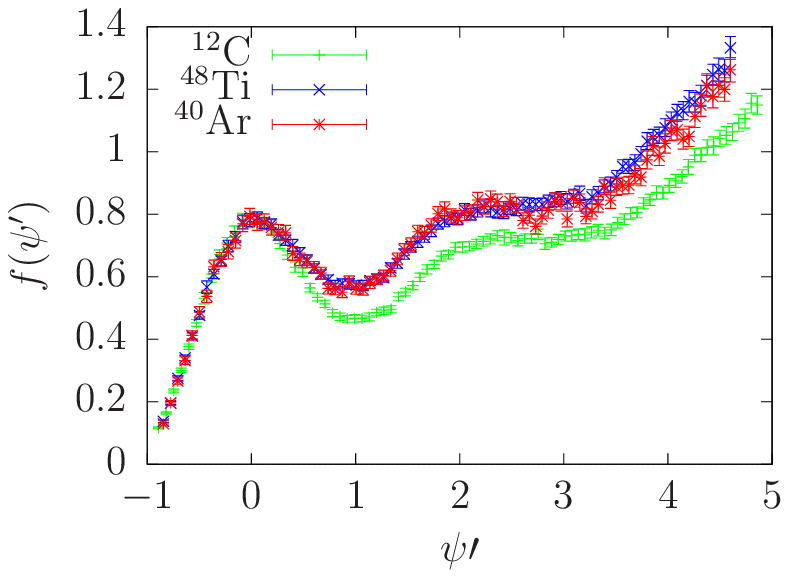}\hspace*{-0.04cm}
\includegraphics[scale=0.675, angle=0]{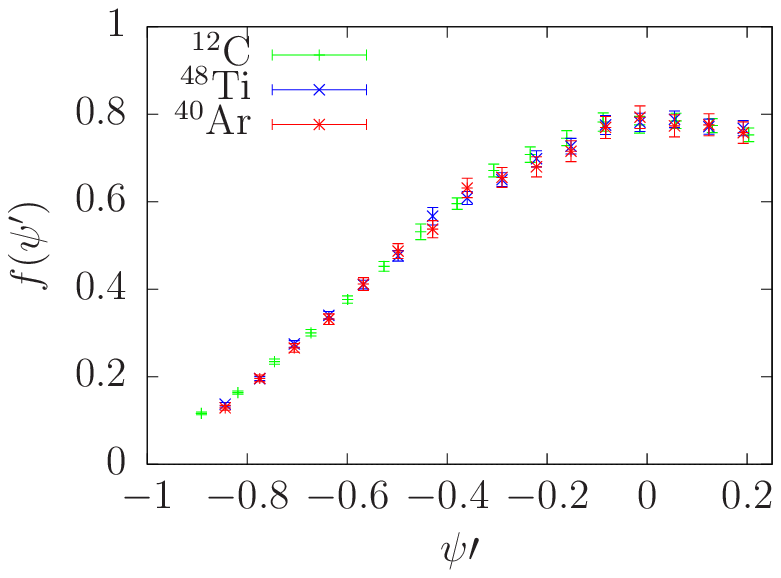}\\
\hspace*{-0.14cm}\includegraphics[scale=0.675, angle=0]{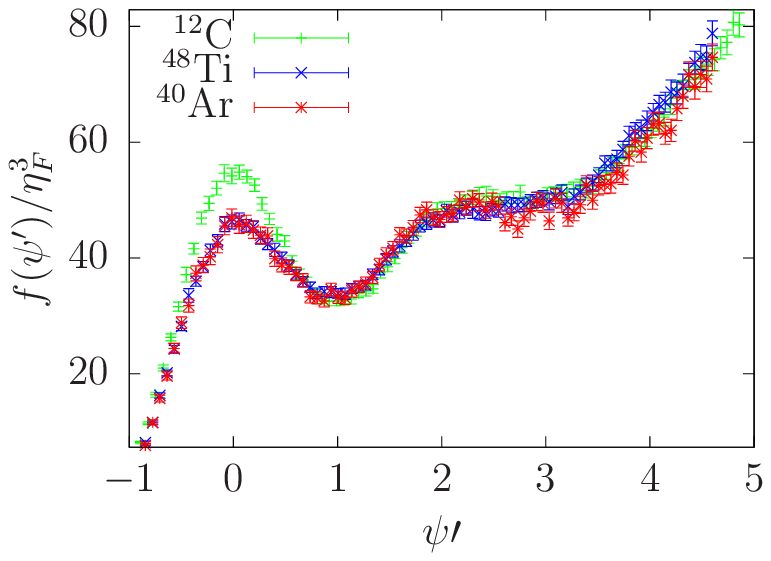}\hspace*{0.1cm}
\includegraphics[scale=0.675, angle=0]{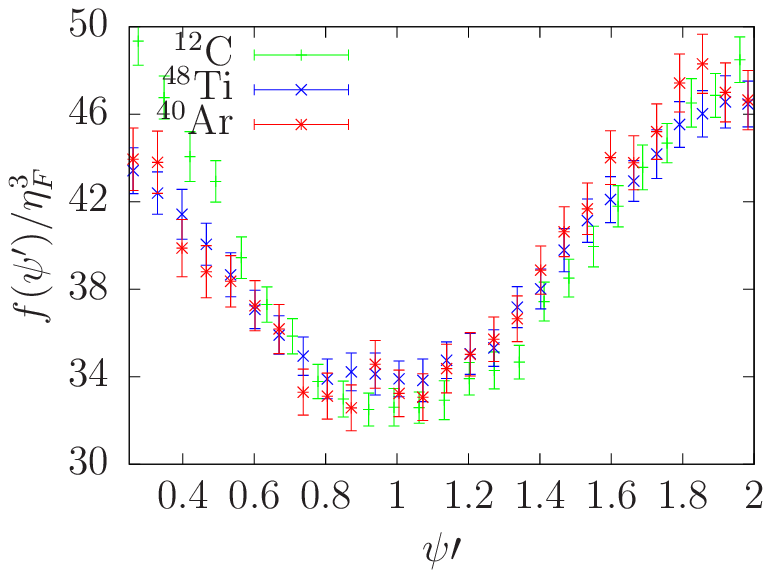}\vspace*{-0.15cm}
	\caption{(Color online)  Top panels: The superscaling function extracted from the JLab data \cite{Dai:2018xhi,Dai:2018gch}. Bottom panels: The superscaling function divided by $\eta_F^3$ extracted from the JLab data \cite{Dai:2018xhi,Dai:2018gch}.}
	\label{fig:fig2}
\end{figure*}

In the top left panel of Fig.~\ref{fig:fig2} we present the superscaling function, as given in Eq.~(\ref{fscaling}), but extracted directly from the experimental data. We show the whole energy spectrum for the three nuclei, C, Ti and Ar. As observed, scaling of second kind only works in the QE peak, then 
it breaks down because non-impulsive contributions (2p2h) and inelastic channels come into play. This was expected from superscaling~\cite{Donnelly99a,Donnelly99b,Maieron:2001it} and was also noticed in  \cite{Dai:2018xhi,Dai:2018gch}. In the region dominated by 2p2h and inelastic contribution, the results for $^{12}$C are significantly below the ones for $^{48}$Ti and $^{40}$Ar which are more in accordance. In order to set how precise scaling of second kind occurs in the QE peak, in the top right panel of Fig.~\ref{fig:fig2} we zoom in to the region at the left of the QE peak (up to $\psi^\prime\simeq 0.25$). Results prove without ambiguity that scaling of second kind is excellent, that is, the superscaling function in this region is universal, being valid up to high precision, for all nuclei. Notice that all data overlap, being aligned in an extremely narrow region. This is consistent with previous analyses. 


The behavior of data with scaling is further investigated in the results shown in bottom panels of Fig.~\ref{fig:fig2}. Here we present the superscaling function $f(\psi^\prime)$ but divided by $\eta_F^3$, where $\eta_F\equiv k_F/m_N$. The panel on the left presents the results for the whole energy spectrum, while the panel on the right is restricted to the dip region, namely, the region where the 2p2h-MEC contributions are more important. As shown, scaling is highly broken in the QE peak (results for carbon are significantly higher), however, results collapse into a single curve within the dip region. This is clearly observed in the right panel where it is shown that all data overlap within their error bars. This
result 
confirms our previous study in \cite{Amaro:2017eah}, where we predicted that 2p2h response scales as $k_F^3$. It is important to point out that the minimum in the cross section shown in Fig.~\ref{fig:fig2} (bottom right panel) corresponds to the maximum in the 2p2h contribution (see Fig.~\ref{fig:fig1}). Although contributions from the QE and inelastic domains also enter in the dip region, and this may at some level break the scaling behavior, results in bottom panels of Fig.~\ref{fig:fig2} strongly reinforce our confidence in the validity of our 2p2h-MEC model, whose predictions are very successfully confirmed by the experimental data. It is also interesting to note that the same type of scaling, i.e, $f(\psi^\prime)/\eta_F^3$, seems to work reasonbly well not only in the dip region but also in the resonance and DIS domains. Further studies on the origin of this behavior are underway.


To conclude, it is noteworthy that SuSAv2-MEC model has proved its capability to describe successfully not only electron but also charged-current neutrino scattering off isospin symmetrical, $N=Z$, nuclei. In the case of electron scattering, the model has been extended to the high inelastic domain providing a good description of data over the whole energy spectrum. In this work, the model is applied for the first time to non-zero isospin nuclei comparing its predictions with the recent data taken at JLab for inclusive electron scattering on carbon, titanium and argon. SuSAv2 makes use of a set of superscaling functions extracted from the predictions provided by the RMF/RPWIA approach. Thus, we have evaluated the RMF/RPWIA scaling functions corresponding to the asymmetric Ti and Ar nuclei, and have compared with the results for $N=Z$ systems, as carbon and calcium. The analysis performed shows unambiguously that scaling of second kind works to high precision which reinforces the validity of one of the main SuSAv2 assumptions. 
Not only are the scaling arguments strengthened, but also the model provides an excellent 
description of the recent JLab data over the whole energy spectrum.
Some small discrepancy at very high energy transfer will be corrected in the next version of the model, which is currently under construction. The good agreement with these data gives us confidence in applying SuSAv2-MEC to neutrino scattering on asymmetric nuclei.

The scaling behavior of the QE and 2p2h responses has been investigated at depth by showing the superscaling function extracted directly from the JLab data. As in previous studies for symmetric nuclei, scaling of second kind works extremely well in the region at the left of the QE peak. On the contrary, the analysis of the dip region, where 2p2h get their largest contributions, proves that data obey a different scaling law, such that the ratio between the 2p2h and the QE responses scales as $k_F^3$. This result, predicted in \cite{Amaro:2017eah}, confirms and strengthens the validity of the 2p2h MEC model.

Finally, we emphasize the importance of scaling arguments applied also to non-zero isospin nuclei, and the successful description within the SuSAv2-MEC framework of the inclusive electron scattering data on C, Ti and Ar recently measured at JLab. This together with the capability of the SuSAv2-MEC approach to describe electron and neutrino interactions on different nuclei, translating sophisticated and demanding microscopic calculations into a straightforward formalism, makes this model a promising candidate to be employed in MonteCarlo event generators (NEUT~\cite{Hayato:2009zz}, GENIE~\cite{Andreopoulos:2009rq} and NuWro~\cite{PhysRevC.86.015505}) used in neutrino oscillation analyses. Accordingly, collaborations with experimental groups at FermiLab (MINER$\nu$A, MicroBooNE) and J-PARC (T2K) are being carried out to implement the SuSAv2-MEC model for electron and neutrino reactions in event generators with rewarding preliminary results~\cite{Megias:2018genie}.

 This work was partially supported by the Spanish Ministerio de Economia y Competitividad and ERDF (European Regional Development Fund) under contracts
FIS2017-88410-P, by the Junta de Andalucia (grant No. FQM160), and part (T.W.D.) by the U.S. Department of Energy under cooperative agreement DE-FC02-94ER40818. M.B.B. and A.D.P. acknowledge support by the INFN under project Iniziativa Specifica MANYBODY and the University of Turin under Project BARM-RILO-17. 
  R.G.J. was supported by Comunidad de Madrid and U.C.M. under the contract No. 2017-T2/TIC-5252.
  G.D.M. acknowledges support from the University of Seville and CEA-Irfu under a Junta de Andalucia fellowship (FQM7632, Proyectos de Excelencia 2011). We are in debt with J.M. Ud\'ias for providing us codes about the RMF model.

%

\bibliography{biblio}

\end{document}